\documentclass{ws-ijmpd}
\usepackage[super,compress]{cite}
\usepackage{psfig}

\begin{document}

\markboth{Bhal Chandra Joshi}
{Pulsar Timing Arrays}

%
\catchline{}{}{}{}{}
%

\title{PULSAR TIMING ARRAYS}

\author{BHAL CHANDRA JOSHI}

\address{National Center for Radio Astrophysics (TIFR), \\
Post Bag No 3, Ganeshkhind, Pune - 411007 (India)\\
bcj@ncra.tifr.res.in}

\maketitle

\begin{history}
\received{Day Month Year}
\revised{Day Month Year}
\end{history}

\begin{abstract}
In the last decade, the use of an ensemble of radio pulsars to 
constrain the characteristic strain caused by a stochastic 
gravitational wave background has advanced the cause of detection 
of very low frequency gravitational waves significantly. This 
electromagnetic means of gravitational wave detection, called 
Pulsar Timing Array(PTA), is reviewed in this article. The principle 
of operation of PTA, the current operating PTA and their status 
is presented along-with a discussion of the main challenges 
in the detection of gravitational waves using PTA.
\end{abstract}


\ccode{PACS numbers:97.60.Jd 97.60.Gb 95.85.Sz}


\section{Introduction}	

A distinguishing feature between the Newtonian 
gravity and General relativity (GR) is the 
existence of gravitational waves, which are implied 
by any relativistic theory of gravitation. Einstein
proposed the notion of gravitational wave (GW) 
formally, soon after formulating GR.\cite{e1918}
As these waves can propagate without any 
significant absorption, they carry information
about large gravitating masses, 
often exclusively, and complementary to that 
provided by electromagnetic waves. Such information 
is important for the development of gravity 
theories and by extension, for a fundamental 
understanding of the origin and
evolution of our universe. 

GW astronomy has 
seen significant development in the last 4 
decades since the first instruments were conceived. 
\cite{w1969} Sensitive terrestrial instruments,  
such as Laser Interferometer Gravitational-Wave 
Observatory (LIGO) and Virgo, are operational 
and space-based instruments, such as 
New Gravitational Wave Observatory (NGO/eLISA)\cite{j+11} 
and ASTRO-GW\cite{n13}, are 
proposed in the future to detect these waves. 
International Pulsar Timing Array (IPTA) is an ambitious
collaborative international effort of the 
entire radio pulsar community to detect gravitational
waves using pulsed electromagnetic radiations 
from pulsars. While LIGO (and advanced LIGO) 
is sensitive to a frequency range between 10 Hz 
to 100 KHz, NGO/eLISA would be sensitive between 
0.1 Hz to few $\mu$Hz. IPTA and its constituent 
Pulsar Timing Arrays (PTAs), owing to their large 
arm length, probe GW in the nano-Hz range (very low 
frequency band\cite{t95,n10} - 300 pHz to 100 
nHz\footnote{http://astrod.wikispaces.com/file/view/GW-classification.pdf}.) 
and are complementary with the terrestrial and 
space based GW detectors. A review of the current
status of GW detection with the PTAs is presented in
this paper.

Pulsars are highly magnetized rotating 
neutron stars emitting a train of narrow 
radio pulses. They are known for their 
remarkable stability  of period, which 
represents the rotation rate of the star. 
The large rotational kinetic energy 
reservoir (10$^{46}$ ergs) of this 20-km
diameter star, with typical mass 1.4  
$M_\odot$, is responsible for 
its remarkable clock like behaviour. 
As clocks play an important role in 
characterizing gravity theories, radio
pulsars have been used to test these 
theories to a remarkable precision.\cite{ksm+06}  
The first detection, albeit an indirect 
one, of the existence of GW was also 
provided by radio pulsars, soon after 
the discovery of the first double 
neutron star (DNS) system, PSR B1913+16.\cite{ht75a,tw82}  
As of today, 9 DNS systems are known\cite{obg+11} apart 
from other binary systems involving radio 
pulsars, which continue to provide new 
tests for gravitational physics. PTA utilizes 
the clock-like mechanism of radio pulsars to 
construct a telescope for the detection of very low
frequency GW.

A brief historical account of PTAs is in order 
here. The idea to use pulsar timing residuals to 
constrain the stochastic GW background (SGWB), 
considering root mean square (rms)  
timing residuals of about 150 $\mu$s over
5 years of data in long  period pulsars such 
as PSRs B1919+21, B1933+16 and B2016+28, was first
proposed   by 
Detweiler\cite{detw79} in 1979 closely following 
a suggestion by Sazhin\cite{saz78} in 1978. 
Hellings \& Downs\cite{hd83} further
analysed data from 4 slow pulsars using NASA's 
deep space network antennas to put a limit on 
SGWB to about 0.01 percent of
the critical/closure density. A similar limit 
from B1237+25 was arrived at by 
Romani \& Taylor\cite{rt83}, who also compared 
it with other methods discussed by 
Zimmerman \& Hellings.\cite{zh80}. Millisecond 
pulsars (MSPs), pulsars with
short period (ms) and low magnetic 
fields, provide much higher period stability 
and precision of measurements of the 
time-of-arrival (TOA) of the radio pulse. Hence, 
MSP timing started soon after the discovery of
the first MSP B1937+21\cite{bkh+82} and 
interesting limits from long term timing of MSPs 
were put by several groups\cite{fb90,ktr94}, 
the most stringent one coming from the latter 
authors. Work of these groups, particularly 
Kaspi et al.\cite{ktr94} also highlighted 
challenges presented by an instrument, which uses 
pulsar clocks, to set a limit or detect GW, namely, 
errors due to ephemeris, Dispersion Measure (DM)
\footnote{Dispersion Measure (pc\,cm$^{-3}$) is 
defined as the integrated column density of 
electrons along the line of sight} variations and so on. 
Both Detweiler and Hellings \& Down\cite{detw79,hd83} 
had proposed using a pair or ensemble of pulsars 
as a means to improve sensitivity not only to 
SGWB, but
also towards individual sources. This was further  
extended by Foster \& Backer\cite{fb90} to propose 
a GW telescope using an array of pulsars.  
Currently, three such experiments implement a 
substantially developed form of this concept and 
are also sharing their data in a major international
effort to address this issue.

Although such a celestial instrument was proposed 
more than 3 decades ago, advances
in instrumentation and our knowledge of systematic 
effects in pulsar timing, based on two decades of 
dedicated sustained timing of a large number 
of pulsars from Jodrell Bank and Parkes radio 
telescope and other smaller scale efforts, have 
accelerated efforts in the last ten years to realise 
such an instrument. A sustained international
collaborative effort of the entire pulsar community, 
particularly in the last 5 years, appears to be 
bringing such an instrument excitingly close to 
an eventual detection. A review of these extensive
efforts is presented in this paper.

The plan of the paper is as follows. The conceptual 
and theoretical framework and the technique used in 
a  PTA is described in Section \ref{concept}. The 
challenges in the detection of the small amplitude 
of GW are discussed in Section \ref{challenges}. A PTA 
requires an ensemble of pulsars. The criteria for the 
source selection and the observing requirements
are presented in Section \ref{reqforpta}. 
The current PTA initiatives are then presented in Section 
\ref{curpta} followed by a review of the status 
of these efforts in Section \ref{status}. The 
concluding section presents some remarks on 
future desirable developments for PTAs.

\section{An Array of Pulsars as a Gravitational Wave Telescope}	
\label{concept}

A typical PTA attempts to detect correlated 
perturbation in flat space-time, caused by a 
passing GW, by measuring systematic deviations 
from an expected rotation model of a radio pulsar 
using the technique of pulsar timing. A brief discussion 
of these ideas is presented here.

\subsection{Pulsar Timing}
\label{pt}

Pulsar timing is an observational technique 
used to determine accurately the rotation 
parameters of a radio pulsar.  Here we briefly review the technique. 
The pulsed emission from a radio pulsar, 
averaged over several thousands of periods, 
is a unique signature of the pulsar\cite{hmt75} 
and is called the average profile (AP) 
of the pulsar. The time-of-arrival 
(TOA) of a fiducial point on the AP 
is computed by cross-correlating 
the observed AP of a pulsar with a noise 
free template constructed from several 
observations of the pulsar. The observed 
TOAs are then transformed to a proper time 
at the solar system barycentre (SSB)
through a chain of transformations, involving 
pulsar position, proper motion, parallax, 
DM, relativistic clock and space corrections in 
the solar system,  and corrected
for arrival at the SSB. The corrected TOAs 
are then compared with those predicted from 
an assumed  model of the pulsar clock 
to obtain timing residuals. The assumed model of the 
pulsar clock involves the rotation 
rate of the pulsar and its higher derivatives, 
and may involve, for binary pulsars, 
transformations similar to those outlined
above from binary centre of mass to pulsar 
frame involving Keplerian, and post-Keplerian 
parameters, if the binary is relativistic. The 
timing residuals are then fitted in a least 
squares fit to improve the parameters of the 
assumed model. If the fitted model describes the pulsar clock
well, one should obtain ``white'' or Gaussian 
distributed residuals at the end of the 
process representing only the measurement
errors. Any systematics or larger than 
expected errors indicate unmodeled parameters.
The techniques and the standard software, 
TEMPO(2), used for this analysis are described in 
Refs. \citen{mantay77,lynsm98,lorkram05,bh86,hem06,ehm06}.   
The perturbation produced in the metric by
a passing gravitational wave is one such 
parameter. Thus, rms timing residual provides 
an upper limit on this perturbation 
and can be used to constrain the amplitude 
of GW.\cite{bh86}

\subsection{Effect of GW on timing residuals}
\label{gwtr}

A passing GW introduces small time varying 
changes in the proper separation between 
two points in space-time. Thus, 
the time of flight of photons of a 
received electromagnetic radiation varies 
introducing a shift in frequency, a fact 
first noted by Estabrook \& Wahlquist.\cite{ew75} 
GW detectors, such as LIGO 
and LISA, use phase difference obtained with  
interferometric measurements, between 
the emitted and the reflected radiation from 
a coherent source, such as a laser, to 
measure these changes in the proper separation. 
On the other hand, PTA employs a natural coherent 
source, namely pulsars, by measuring 
its apparent pulse frequency. The fractional 
difference in the apparent pulse frequency, $\nu$(t), 
and the pulse frequency emitted by the pulsar, 
$\nu_0$, is given by the following expression 
for a monochromatic GW, propagating in z 
direction in a coordinate system with earth 
at origin\cite{bh86}
\begin{equation}
\frac{(\nu(t) -\nu_0)}{\nu_0} = \frac{1}{2}(1 - \cos(\theta)) \cos(2 \phi) [
  h(t) - h(t - d/c - d \cos(\theta)/c)]
\label{appfreq}
\end{equation}
where, h is the amplitude of the wave. 
The earth-pulsar sight-line 
is at an angle $\theta$ in the x-z plane and the 
principal polarisation of the wave makes 
an angle $\phi$ with the x-axis. The left hand 
side of Eq.~\ref{appfreq} is equivalent 
to the derivative of the timing residual, thus relating 
the timing residuals to the amplitude of the GW. 
In particular, the spectral density of the residuals 
is identical to that of the amplitude of the wave\cite{detw79} 
and can be obtained from the auto-correlation of 
the residuals.  

The duration used to obtain the expected mean square 
residuals in the above auto-correlation sets the frequency 
range of the GWs to which the PTA is sensitive. Typical 
durations quoted are in the range of 10$^7$ to 10$^8$ s, 
corresponding to arm lengths in excess of 10$^{15}$ km and 
frequencies of 1 to 100 nHz. This distinguishes 
PTA from other GW detectors, where the electromagnetic 
radiation travels a very small fraction of the GW 
wavelength. This range of frequency is also 
complementary to other GW detection experiments.

\subsection{Upper Limits on SGWB using PTA}
\label{SGWBPTAcon}
 
The SGWB can be specified 
in terms of three different quantities of interest, 
which are related to each other and can be estimated 
from the timing residual measurements. Much of the earlier 
work used the estimates of rms timing residuals 
from individual pulsars to obtain an upper limit 
on the energy density of the GW as a fraction 
of the critical density in an FRW (Friedmann-
Robertson-Walker) universe given by 
\cite{detw79,ar99,m00,jb03} 
\begin{equation}
\Omega_{gw}(f) = \frac{8 \pi G}{3 H_0^2}\frac{d\rho_{gw}}{d \log f}
\label{omegagw}
\end{equation}
where, $H_0 = $ 100 $h_0$ km s$^{-1}$ is 
Hubble's constant with $h_0$ expressing the uncertainty 
in the Hubble constant (typically 0.5 $< h_0 <$ 0.65\cite{m00}).  
The dimensionless energy density, $\Omega_{gw}$(f), has an 
uncertainty due to $h_0$. The ensemble average, over all 
directions and sources, of Fourier component of the metric 
perturbation is defined as the spectral density, $S_h$(f). 
It can be estimated from the spectral density of the timing 
residuals. Since  $S_h$(f) has the dimensions of inverse 
frequency, a dimensionless quantity, $h_c$(f), 
characteristic of strain amplitude, is often used and is 
related to $S_h$(f) and  $\Omega_{gw}$(f) as below
\begin{equation}\label{charstrain}
\begin{split}
h_c^2(f) & = 2 f S_h(f) \\
& = \frac{ 3 H_0^2 \Omega_{gw}(f)}{2 \pi^2 f^2 } 
\end{split}
\end{equation}
The characteristic strain is related to the spectra of the timing 
residuals \cite{jhs+06}
\begin{equation}
h_c^2(f) = 12 \pi^2 f^3 R(f)
\label{hinres}
\end{equation}
where, R(f), is the power spectrum of the timing residuals. The 
variance of arrival time residuals, $\sigma^2$, is related to R(f) by 
the following expression
\begin{equation}
\sigma^2 = \int_0^\infty  R(f) df
\label{sigres}
\end{equation}

In general, there are many contributions,
other than that due to GW, to the  noise 
seen in the timing residuals (See Section \ref{challenges}). 
Hence, the estimate arrived with Eqs. \ref{hinres} 
and \ref{sigres} can 
only provide an upper limit to the characteristic 
strain, $h_c$. Thus, individual pulsars with small timing 
residuals and ``white'' residuals provide the best 
upper limits. Much of the earlier work on the 
detection of SGWB used this 
method to estimate an upper limit by averaging 
the measurements  of $\sigma^2$ for a sample of 
pulsars, assuming each to be an independent 
measurement. 

The change in apparent frequency, given in Eq. \ref{appfreq}, 
incorporates information about GW amplitudes
at two distinct space and time coordinates. The 
first term incorporates information about h at 
the time and place of reception (at Earth) and 
the second at the pulsar. The first term will be 
correlated for many pulsars, whereas 
the factors contributing to the timing noise 
(See \ref{challenges}) of two pulsars are 
generally uncorrelated. Monitoring of timing 
residuals in this coherent fashion from 
a large sample of pulsars is what constitutes 
a PTA. The aim of  PTAs is to detect this 
correlation as a function of  
angular separation between pairs of 
pulsars in the PTA. 

\begin{figure}[pt]
\centerline{\psfig{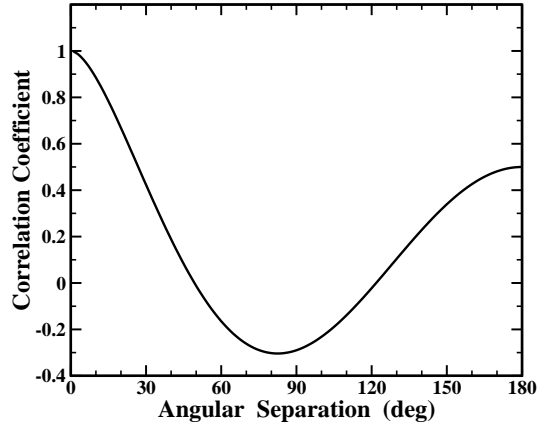}}
\vspace*{8pt}
\caption{Expected correlation between timing residuals 
for a pair of pulsars as a function of separation angle\cite{hd83}}
\label{hd83plot}
\end{figure}

The form of 
this correlation, $\zeta(\theta_{ab})$
\footnote{The scaling factor used in 
this definition for $\zeta$ varies in the 
literature. We have adopted the definition from 
Ref. \citen{hd83}}, 
is given by \cite{hd83,jhl+05,bd08}
\begin{equation}
\zeta(\theta_{ab}) = \frac{( 1 - \cos(\theta_{ab}))}{2} 
  \ln(\frac{(1 - \cos(\theta_{ab}))}{2}) 
  - \frac{1}{6} \frac{( 1 - \cos(\theta_{ab}))}{2} + \frac{1}{3}
\label{hd83corr}
\end{equation}
which is plotted in Fig. \ref{hd83plot}. The angular 
separation between the pair of pulsars is $\theta_{ab}$. Timing 
observations of a sample of pulsars can be analysed 
to look for a correlation, which provides a constraint 
on the amplitude of the GW spectrum. The characteristic 
strain spectrum for different models of SGWB can be 
generically written as \cite{jhs+06}
\begin{equation}
h_c(f) = A f^\alpha
\label{specthc}
\end{equation}

Thus, a limit can be obtained in principle by a fit 
to correlations in the timing residuals of several 
pairs of pulsars with varying  angular 
separations\cite{dfg+12}.

\subsection{Models of SGWB and characteristic strain spectrum}
\label{SGWBMOD}

While ground and space-based detectors are sensitive 
to GW generated by sources, such as supernovae, spinning 
neutron stars, in-spirals of extreme mass binary systems, 
coalescing binaries and binaries with neutron stars and 
black holes and super-massive black hole binaries (SMBH), 
the most probable sources for PTA are SMBH. 
Galaxy mergers during the history of universe 
can lead to the central SMBH in these galaxies 
to become gravitationally bound as a SMBH binary 
and generate GW during its late time in-spiral. 
A PTA can potentially detect GW from such 
SMBH with orbital periods of 10$^8$ s. 

A large number of SMBH are predicted by the theories of 
formation and evolution of SMBH. The most 
likely source for PTAs is the SGWB resulting 
from an incoherent addition of GW generated 
by a number of such SMBH systems\cite{rr95,jb03}. 
The characteristics strain spectrum for 
such a background is given by Eq. \ref{specthc}, 
with the power law index, $\alpha$ = -2/3\cite{jb03}.

Alternatively, PTAs are sensitive enough to 
detect SGWB due to other models. One such 
alternative is the quantum-mechanical generation 
of cosmological perturbations\cite{g05}. The 
power law index, $\alpha$, in the Eq. \ref{specthc} 
is -1 corresponding to the parameter $\beta = $ 
-2 characterising the power law evolution 
of the cosmological scale factor, $a$, with the cosmic time 
(See Eq. 2 of Ref. \citen{g05}). Another alternative is SGWB 
formed by GW emitted by a network of cosmic 
strings\cite{dv05}. The index in this case is -7/6. 
Therefore, PTA results can put meaningful 
constraints on these models.

\section{Factors affecting PTA precision}
\label{challenges}

Although more than 2000 pulsars have 
been discovered till date, it is difficult 
to obtain an ideal ensemble of pulsars, 
which satisfy all the requirements 
as listed in Section \ref{reqpta}, to form a 
high precision PTA. Apart from the 
pulsars being typically weak radio 
sources, the pulsed emission from pulsar 
exhibits a variety of instabilities, 
which makes extracting extremely tiny 
systematic deviations, caused by GW 
perturbation of metric, very challenging. These 
factors are discussed in this section.

\subsection{Rotational instabilities}
\label{rotinst}

Although pulsars broadly exhibit clock-like 
stability, their rotation is affected by small 
rotational irregularities. Many young pulsars 
(characteristic age, $\tau_c <$ 10$^6$ yr)\footnote{estimated 
from measurements of periods and period 
derivatives assuming dipole radiation from a 
pulsar} show sudden increase in their rotation 
rates, known as glitches\cite{sl96,klgj03,el+11}. 
The frequency change at the glitch and its post-glitch 
recovery complicates timing analysis. Moreover, 
these events occur randomly. Hence, such pulsars 
are excluded from PTAs. 

None of older pulsars ($\tau_c >$ 10$^8$) have been 
reported to glitch and these tend to have periods 
larger than few 100 ms. While such pulsars 
were previously used to put constraints on 
SGWB\cite{hd83}, the longer observations 
required for these pulsars to form a 
stable AP and changes in AP due to a 
variety of effects discussed later preclude their 
inclusion in the sample of pulsars for the PTA. 
Another class of old pulsars not showing glitches 
are the millisecond pulsars (MSPs), which are 
pulsars with low surface magnetic fields 
(B $<$ 10$^{10}$ G). The short periods 
help in obtaining a stable AP in short observations. 
Hence, MSPs are ideal candidates for a PTA sample.

Many MSPs  show slow wander of rotation rate over 
intervals ranging from months to years. 
This phenomena, called timing noise, is also not 
predictable\cite{hlk+10}.  
Timing noise is ``red'' in character 
and is similar to the expected GW signature and 
presents a significant constraint\cite{cs10}. 
Recently, quasi-periodic structures  in timing residuals 
have been reported suggesting an origin intrinsic 
to the neutron star\cite{lhk+10}  (See Section \ref{magneto}). 
Hence, a systematic characterization of  a 
much larger sample of MSPs, than being used in the 
present day PTAs, is required to choose the most ideal 
rotators. 

\subsection{Signal-to-noise considerations}
\label{snr}

The TOA precision depends on the radiometer noise 
(and signal-to-noise ratio; S/N)\cite{t92}. Hence, it 
is important to include strong pulsars in a PTA. 
On the other hand, radio pulsars are weak 
radio sources (typical flux density 
10 mJy; 1 mJy = 10$^{-29}$ W/m$^2$/Hz) and this 
restricts the number of usable pulsars. Nearby MSPs are 
usually chosen for most PTAs. While the 
higher sky background leads to higher radiometer 
noise at low radio frequencies, pulsars are stronger 
at these frequencies due to their steep spectra. 
Thus, the best trade-off is achieved at frequencies 
$>$ 1 GHz.

\subsection{Propagation effects}
\label{propeff}

\begin{figure}[pt]
\centerline{\psfig{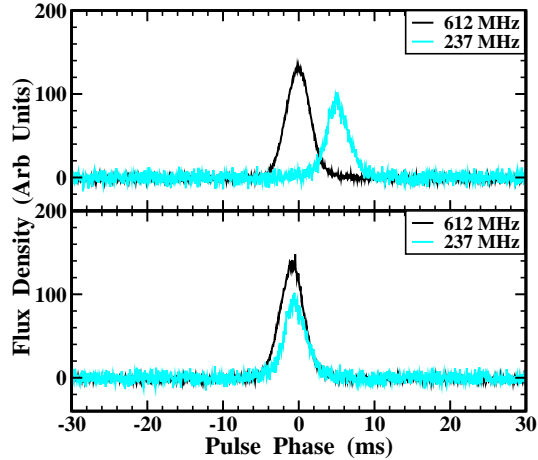}}
\vspace*{8pt}
\caption{The TOA delay due to an error of 0.08 pc\,cm$^{-3}$  in 
assumed DM. Each panel shows the pulse obtained 
at 613 MHz and 237 MHz respectively with the 
GMRT. The upper panel shows the averaged 
pulse at the two frequencies with an assumed DM of 
35.665 pc\,cm$^{-3}$, while the lower panel shows that 
with the estimated DM (35.740 pc\,cm$^{-3}$). 
Figure reproduced from Ref. \citen{jr06}}
\label{dmerr}
\end{figure}

\begin{figure}[pb]
\centerline{\psfig{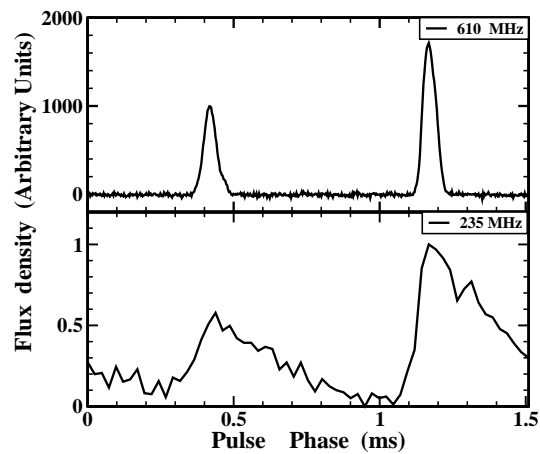}}
\vspace*{8pt}
\caption{Scatter-broadening at low radio frequencies. 
Both panels show coherently dedispersed AP of MSP PSR B1937+21. 
The upper panel shows the AP  at 610 MHz, whereas the lower 
panel shows the AP at 235 MHz, which is clearly scatter-broadened. 
Figure reproduced from Ref. \citen{jr06}}
\label{1937scatbr}
\end{figure}

The deviation in pulsar clock are 
measured using the pulsed radio signal from 
the pulsar, which propagates through the 
inter-stellar medium (ISM). Radio emission is 
dispersed and scattered by the electrons 
in the ISM and encounters multi-path propagation. 
This smears the pulse, reduces the effective time 
resolution and S/N and also causes a frequency 
dependent delay in TOA. A combination of 
all these effects, due to the inhomogeneities in ISM,  
causes variability in pulse TOA, compromising the 
precision of a PTA. An illustration of these 
effects is shown in Fig. \ref{dmerr} and 
\ref{1937scatbr}\cite{jr06}. 

As pulsars are fast moving objects\cite{ll94}, 
observations of pulsars encounter different cross 
sections of ISM on different epochs. Fluctuation of 
electron density over these sections causes the 
dispersive delay in TOA of pulsars to vary. DM 
variations of the order of 0.0001 pc\,cm$^{-3}$ have 
been established in many past studies
\cite{ktr94,agm+05,yhc+07a,yhc+07b,kcs+12,mhb+12}, 
which correspond to variations 
in dispersive delay in TOA of the order of 212 ns 
at 1400 MHz. In addition, the line of sight to a 
few PTA pulsars pass close to Sun at specific 
epochs in a year, causing an additional dispersive 
delay due to the solar wind of the order of 
1 $\mu$s within 7$^o$ of the Sun\cite{yhc+07a,yhc+07b}.

Appropriate corrections, estimated from observations 
at two or more frequencies within a few days of timing 
observations,  for such variations are included in the 
timing model. The usual PTAs do not carry out such 
multi-frequency observations simultaneously, but 
the comparison of the results of such 
non-simultaneous observations with those from 
simultaneous multi-frequency observations, 
in progress at the GMRT, shows that the estimation 
uncertainties are almost identical\cite{kg+12}.

The dispersion of the radio pulse also reduces 
the time resolution for the pulsar observed 
with the traditional filterbank instruments 
due to a finite channel bandwidth. Hence, PTAs 
utilize instruments implementing coherent 
dedispersion in either hardware  or 
software, which allow even better estimation 
of DM variations.

Multi-path propagation of the pulsar signal 
leads to variable broadening of AP (Fig. \ref{1937scatbr}), 
particularly at frequencies below 400 MHz. As 
this is not easily correctable, timing 
observations are usually carried out above 1 GHz, 
where this effect is negligible for low DM 
pulsars due to steep ($\sim \nu^{-4}$) frequency 
dependence of scatter-broadening.

Lastly, intensity variations, usually caused by 
Diffractive Inter-stellar scintillation (DISS) and 
Refractive Inter-stellar scintillation (RISS), 
affect the TOA precision. The former can be 
addressed by employing large bandwidths compared 
to scintle size, which is also desirable for 
higher S/N.

\subsection{Intrinsic changes in pulsed emission}
\label{magneto}

The pulsed emission from a pulsar is the probe of 
its clock mechanism provided the pulse (AP) is stable 
in its shape and arrival phase. Although this was  
broadly established in the early days of pulsar 
astronomy\cite{hmt75}, observations over last 
four decades have shown that many pulsars switch 
between two or more stable APs\cite{l71}. This 
phenomenon is called profile mode-change. 
Recent studies indicate that profile mode-changes 
may be related to/or accompanied by changes in rotational 
parameters of a pulsar. Intermittent pulsars, 
such as PSRs B1931+24, J1841$-$0500, J1832+0029, 
J1839+1506, exhibit extended periods (several 
hundred days) with no pulsed emission and 
different spin-down rates for their ON and 
OFF epochs\cite{klo+06,crs+12,llm+12,sjm+12}. 
A sample of 17 pulsars with profile 
changes corresponding to changes in 
spin-down rate was reported recently 
suggesting the cause to be switching between 
different magnetospheric states\cite{lhk+10}. 
Lyne et al.\cite{lhk+10} argued that the 
timing noise in MSPs can be explained 
by slow profile changes and claimed that 
it may be possible to correct for timing 
noise. In any case, the effect of the 
magnetospheric processes on the TOA 
precision poses a challenge, quite distinct 
from the processes related to the internal 
structure of neutron star, as discussed 
in Section \ref{rotinst}, and motivates a better 
understanding of these processes to account 
for their contribution to timing residuals. 

\subsection{Errors in timing technique}

In addition to the above effects, the timing 
residuals may have contributions from the clock errors in 
the atomic clocks used for obtaining the topocentric 
TOA and errors in the solar system ephemeris. 
See Ref. \citen{fb90} for a detailed discussion 
on these errors. Both the ephemeris and  
the atomic standards are being steadily improved.

\section{Requirements for a PTA experiment}
\label{reqforpta}

The effects discussed in the last section influence 
the choice of pulsars and the observing instrumentation 
best suited for a high precision PTA. These requirements are 
summarised below. 

\subsection{Observing requirements}

The appropriate instrumentation and observing strategy required 
for a typical PTA are

\begin{itemlist}
   \item  A telescope or combination of telescopes with large 
          collecting area
   \item  Wide-band receivers, preferably with a bandwidth $>$ 1$-$2 
          GHz
   \item  Observing frequency between 1 to 3 GHz for the best trade-off 
          between the influence of sky-background, steep pulsar 
          spectra and the propagation effects
   \item  Simultaneous or near-simultaneous observations at 2 or more 
          frequencies to calibrate DM changes, with one frequency 
          preferably below 400 MHz
   \item  Full polarimetric observations
   \item  High time resolution receivers, employing coherent 
          dedispersion, to eliminate the dispersion smearing 
          of the pulse.
\end{itemlist}

The frequent monitoring of a large sample of pulsars 
using the most sensitive telescopes requires a large 
amount of observing time on these, usually 
over-subscribed telescopes. PTAs in operations 
currently try to share this burden, which was 
also a prime motivation for forming the International 
Pulsar Timing Array (IPTA). Use of different observing 
system introduces sometimes epoch dependent variable 
fixed pipeline delays and clock errors. Thus, an important 
requirement for instrumentation is to periodically 
calibrate such delays.

\subsection{Sample selection}
\label{reqpta}

The requirements for the ensemble of pulsars 
comprising an effective PTA are as follows

\begin{itemlist}
   \item  Pulsars with high signal-to-noise ratio (S/N) 
          pulsed emission
   \item  Pulsars with stable AP
   \item  Pulsars with narrow pulses
   \item  Field MSP or MSPs in wide binaries are preferred over 
          tight binaries to avoid relativistic effects in timing
   \item  Pulsars on line  of sights without  unusual  ISM structures
   \item  Nearby pulsars are preferred over more distant ones
   \item  Pulsars with exceptionally  high rotational  stability
   \item  A reasonable  distribution of  pulsars  mimicking the
          pairwise angular separation required for 
          $\zeta(\theta_{ab})$ given by Eq. \ref{hd83corr}
\end{itemlist}

The first six requirements have important implications 
for the precision of detection of a GW signature and are 
the consequences of the timing technique itself. The last two 
requirements are dictated by SGWB detection technique. 

The short periods of MSPs combined with sharp pulses 
allows averaging several thousands of their pulses. 
MSPs also have very stable rotation and a more uniform 
distribution across the sky as they form a relatively 
local population. Hence, all PTA samples use MSPs as 
their ensemble. 

Although more than 150 MSPs are known, the above criteria 
reduces the list of suitable pulsars to about 80. While 
this ensemble is useful, it is not ideal as most of these 
show systematics in timing residuals, such as timing noise. 
Currently, only PSR J1713+0747  
shows white residuals over 5 years, with post-fit 
rms residuals of better than 
200 ns\cite{dfg+12,mhb+12}. More such pulsars 
are required to be discovered to make an 
effective PTA.

\section{Pulsar Timing Arrays}
\label{curpta}

There are three currently active pulsar timing 
arrays - the Parkes Pulsar Timing Array (PPTA),  
the North American Nanohertz Observatory for 
Gravitational Waves (NANOGrav) and the 
European Pulsar Timing Array(EPTA). A brief description 
of each of these PTAs highlighting the instrumentation 
in use is presented in this section. The details 
can be found in Refs. \citen{mhb+12,dfg+12,hlj+11}.

\subsection{The Parkes Pulsar Timing Array (PPTA)}

PPTA, which started operating in 2004, 
was initiated by R. N. Manchester and 
is a collaboration between Australia Telescope National 
facility (ATNF), Swinburne University of Technology 
and several other Australian and international  
institutions. It uses 64-m Parkes radio 
telescope to time 20 MSPs every 2 to 3 weeks. The Parkes 
radio telescope can observe MSPs below a declination of 
+ 25$^0$ and the entire southern Galactic plane, 
allowing it to observe a large sample of Galactic MSPs.
Observations are carried out using 20-cm, 10-cm and 
50-cm receivers, the latter two frequencies coming 
from a dual-band coaxial receiver. All these are 
wide-band receivers. Wide-band backends, such as 
Parkes Digital Filterbank systems (PDFB1/PDFB2/PDFB3/
PDFB4) and baseband recording systems with 
coherent dedispersion, such as Caltech-Parkes-
Swinburne-Recorder (CPSR2) and APSR are employed for 
the observations. See Ref. 
\citen{mhb+12} and references therein for details 
of the observing system, instrumentation used, 
data reduction software and off-line processing 
pipeline.

\subsection{The North American Nanohertz Observatory for Gravitational Waves (NANOGrav)}

NANOgrav is a collaboration between National 
Radio Astronomy Observatories (NRAO) and 
several other US universities. It primarily uses 
300-m Arecibo radio telescope and 100-m Green Bank 
Telescope(GBT) to time 35 MSPs once every twenty days 
at two out of 327/400/800/1400/2350 MHz wavebands. 
The two frequency observations are not simultaneous, 
but are carried out with a gap ranging from 
1 hour to 7 days. Both Arecibo and GBT are very 
sensitive telescope, thus providing high S/N 
data in 15 to 45 minutes observations. 
Arecibo typically observes 19 pulsars and 
18 pulsars are monitored by GBT. The whole 
ensemble is covered in 18 hours per epoch. Wide-band 
backends, such as Astronomical Signal Processor (ASP) 
or Green Bank Astronomical Signal Processor (GASP), 
are used to provide full polar coherently dedispersed 
data. See Ref. \citen{dfg+12} and references therein 
for details  
of the observing system, instrumentation used and 
timing analysis.

\subsection{The European Pulsar Timing Array(EPTA)}

EPTA is the most recently started PTA program. 
It is a collaboration between several European 
institutions, such as Max-Planck Institute for Radio 
Astronomy, Jodrell Bank Centre for Astrophysics, ASTRON 
and so on. It uses four major European radio telescopes 
$-$ 70-m Lovell telescope at Jodrell Bank, 100-m 
Effelsberg telescope at Bonn, 
Westerbok Synthesis Radio Telescope and Nancy 
radio telescope. Sardinia Radio Telescope is likely 
to be added in future. There are plans to combine 
all these telescopes coherently to form a Large 
European Array for Pulsars (LEAP), which is likely 
to provide a telescope equivalent to 200-m single dish.
See Refs. \citen{fhb+10,hlj+11} and references 
therein for details.

\subsection{International Pulsar Timing Array}

PTA requires frequent monitoring of pulsars at 
multiple radio frequencies. This adds up to a large 
observing time requirement from the participating 
radio telescopes. The pulsar community 
has therefore come together, in a unique initiative, 
to divide the required observing time across all 
participating telescopes and share the data. This 
consortium is called International Pulsar Timing Array 
(IPTA)\footnote{More information can be found at 
http://www.ipta4gw.org/}, which is a collaboration of the three existing 
PTAs - PPTA, NANOgrav and EPTA, and encompasses 
almost the entire pulsar community. 

Apart from the division of observing time, such an 
international effort has other advantages 
and spin-offs. Firstly, this allows a fairly uniform coverage 
of MSPs in the north and south hemispheres. Secondly, the 
participating PTAs will have overlaps in their sample of 
MSPs, which provides consistency and quality checks on 
the data. Thirdly, the data from all these telescopes 
are brought together in public domain, which can lead to 
high quality data sets after proper assessment. Fourthly, this 
allows developing, comparing and contrasting alternative 
analysis techniques for data analysis through regular 
public data challenges. Lastly, such an effort will be 
very useful in developing the gravitational wave community.  

IPTA monitors an ensemble of about 30 MSPs with a view to 
detect SGWB, mainly from super massive black hole binaries, 
with a detector arm length of light years in the nanohertz 
frequency range. Thus, it is complementary to ground (space) 
based detectors with typical arm length of few (millions) 
km and much higher frequency range. 


\section{Current Status of GW detection by PTAs}
\label{status}

The currently operating PTAs are on the threshold 
of detection of SGWB and are already putting 
stringent limits to the expected strain 
from theoretical calculations. A large part of PTA 
effort has been expended in finding the ``ideal'' 
clocks for this challenging and difficult experiment. 
With more than 20000 TOAs on all PTA pulsars, timing residuals of the 
order of 30 ns have been obtained for PSRs J1713+0747 and 
J1909-3744\cite{dfg+12}, which show almost ``white'' timing residuals, 
demonstrating that such high timing precision is in principle 
achievable. These results from NANOGrav are consistent 
with those from PPTA\cite{mhb+12}, which also lists 
PSR J0437$-$4715 as a useful clock. 

As discussed in Section \ref{SGWBPTAcon}, upper limits on 
characteristic strain, $h_c$, can be estimated from these 
individual measurements. These results from the PTAs 
are shown in Table~ 
\ref{spsrlim}.  It is evident that the limits have 
improved substantially and  are already comparable 
with the expected characteristic strain due to such SGWB.

\begin{table}[ph]
\tbl{Summary of limits on h$_c$ from estimates on individual pulsars, 
assuming $\alpha$=$-$2/3}
{\begin{tabular}{lcl} \toprule
Source & h$_c$(per year) & Reference \\ \colrule
B1133+16   & $<$ 9 $\times$ 10$^{-13}$ & Ref. \citen{hd83} \\
B1237+25   & & \\
B1604$-$00 & & \\
B2045$-$16 & & \\ \colrule
B1855+09   & $<$ 2 $\times$ 10$^{-14}$ & Ref. \citen{ktr94} \\
B1937+21   & & \\ \colrule
B1855+09   & $<$ 1.1 $\times$ 10$^{-14}$ & Ref. \citen{jhs+06} \\ \colrule
J1713+0747 & $<$ 1.1 $\times$ 10$^{-14}$ & Ref. \citen{dfg+12} \\
J1909$-$3744&$<$ 3.9 $\times$ 10$^{-14}$ & Ref. \citen{dfg+12} \\
B1855+09   & $<$ 1.3 $\times$ 10$^{-14}$ & Ref. \citen{dfg+12} \\ \colrule
J0437$-$4715&$<$10$^{-15}$  & Ref. \citen{mhb+12} \\
J1713+0747 & $<$10$^{-14}$  & Ref. \citen{mhb+12} \\
J1909$-$3744&$<$10$^{-14}$  & Ref. \citen{mhb+12} \\ \botrule
\end{tabular} \label{spsrlim}}
\end{table}

As discussed in Section \ref{SGWBPTAcon}, upper limits can 
also be obtained by examining the correlations in the 
timing residuals of several pairs of pulsars with 
varying  angular separations. The best fit for the NANOGrav 
data to the Hellings \& Downs correlation curve (Eq. 
\ref{hd83corr}) is consistent with no detectable SGWB 
and translates to a 2$-\sigma$ upper limit of 7.2 $\times$ 
10$^{-15}$, 4.1 $\times$ 10$^{-15}$ and 3.0 $\times$ 10$^{-15}$ 
assuming $\alpha = -$2/3,$-$1 and  $-$7/6 respectively 
for different models of SGWB\cite{dfg+12}. Likewise, 
PPTA data are also consistent with no detectable SGWB, 
but with a more conservative upper limit of 6.0 $\times$ 
10$^{-15}$\cite{ych+11}.

\section{Conclusions and summary}

The current sensitivity of PTAs is tantalizingly 
close to the expected SGWB although no signature 
of such a background is evident from the data 
so far. The best ``clocks'' are PSR J0437$-$4715, J1713+0747 
and  J1909$-$3744. The most stringent limits 
in both methods employed in PTA are dominated 
by these pulsars. The real challenge for the 
PTA is in obtaining a larger sample of such ideal 
``clocks'' sampling the correlation function in 
Fig. \ref{hd83plot} uniformly. This motivates 
new surveys for finding new MSPs. New high-energy 
telescopes such as Fermi-LAT have already revealed 
a much larger population of MSPs and deeper and high 
time resolution radio surveys, such as HTRU survey, 
are underway in both the hemispheres. Another 
challenge is to rapidly characterise the new 
MSPs to be included in the future PTAs. Last, 
but not the least, a better understanding 
of magnetospheric physics of radio pulsars 
is required to model the variable torques 
on the neutron star in a bid to correct 
the covariant ``red noise'' component in the 
residuals contributed by such torques. While the 
SGWB detection still seems some years away, 
these efforts will certainly help advance 
the understanding of physics of radio pulsars 
as well as their environments.

\section*{Acknowledgments}

The author thanks the organisers of ASTROD 5 symposium 
for stimulating this review and R. N. Manchester 
for useful discussions. The author also thanks M. 
McLaughlin, M. Kramer and B. Stappers for a discussion 
on NANOgrav and EPTA and the recent  results from these 
PTAs.

\bibliographystyle{ws-ijmpd}
\bibliography{ptabib}

\begin{thebibliography}{10}

\bibitem{e1918}
A.~Einstein, { About {G}ravitational waves}, in {\em {P}roceedings of
  {P}russian {A}cademy of {S}ciences\/},  (1918), pp. 154--167.

\bibitem{w1969}
J.~Weber, {\em Phys. Rev. Lett.} {\bf 22}  (1969) 1320.

\bibitem{j+11}
O.~Jennrich~et al., {\em ESA/SRE}   (2011) 19.

\bibitem{n13}
W.-T. Ni, {\em Int. J. Mod. Phys. D}   (2013) 1341004,
  arXiv:astro-ph/1212.2861.

\bibitem{t95}
K.~S. Thorne, { Gravitational waves}, in {\em Particle and Nuclear Astrophysics
  and Cosmology in the Next Millenium\/},  eds. E.~Kolb and R.~Peccei. (World
  Scientific, Singapore, 1995), pp. 160--184.

\bibitem{n10}
W.-T. Ni, {\em Mod. Phys. Lett. A} {\bf 25}  (2010) 922.

\bibitem{ksm+06}
M.~Kramer, I.~H. Stairs, R.~N. Manchester, M.~A. McLaughlin, A.~G. Lyne, R.~D.
  Ferdman, M.~Burgay, D.~R. Lorimer, A.~Possenti, N.~D'Amico, J.~M. Sarkissian,
  G.~B. Hobbs, J.~E. Reynolds, P.~C.~C. Freire and F.~Camilo, {\em Science}
  {\bf 314}  (2006) 97.

\bibitem{ht75a}
R.~A. Hulse and J.~H. Taylor, {\em Astrophys. J.} {\bf 195}  (1975) L51.

\bibitem{tw82}
J.~H. Taylor and J.~M. Weisberg, {\em Astrophys. J.} {\bf 253}  (1982) 908.

\bibitem{obg+11}
S.~Oslowski1, T.~Bulik, D.~Gondek-Rosinska and K.~Belczynski, {\em Mon. Not. R.
  Astron. Soc.} {\bf 413}  (2011) 461.

\bibitem{detw79}
S.~Detweiler, {\em Astrophys. J.} {\bf 234}  (1979) 1100.

\bibitem{saz78}
M.~V. Sazhin, {\em Soviet Astr.} {\bf 22}  (1978) 36.

\bibitem{hd83}
R.~W. Hellings and G.~S. Downs, {\em Astrophys. J. Lett.} {\bf 265}  (1983)
  L39.

\bibitem{rt83}
R.~W. Romani and J.~H. Taylor, {\em Astrophys. J. Lett.} {\bf 265}  (1983) L35.

\bibitem{zh80}
R.~L. Zimmerman and R.~W. Hellings, {\em Astrophys. J.} {\bf 241}  (1980) 475.

\bibitem{bkh+82}
D.~C. Backer, S.~R. Kulkarni, C.~Heiles, M.~M. Davis and W.~M. Goss, {\em
  Nature} {\bf 300}  (1982) 615.

\bibitem{fb90}
R.~S. Foster and D.~C. Backer, {\em Astrophys. J.} {\bf 361}  (1990) 300.

\bibitem{ktr94}
V.~M. Kaspi, J.~H. Taylor and M.~F. Ryba, {\em Astrophys. J.} {\bf 428}  (1994)
  713.

\bibitem{hmt75}
D.~J. Helfand, R.~N. Manchester and J.~H. Taylor, {\em Astrophys. J.} {\bf 198}
   (1975) 661.

\bibitem{mantay77}
R.~N. Manchester and J.~H. Taylor, {\em Pulsars} (Freeman, San Fransisco,
  1977).

\bibitem{lynsm98}
A.~G. Lyne and F.~G. Smith, {\em Pulsar Astronomy}, 2nd edition edn. (Cambridge
  University Press, Cambridge, 1998).

\bibitem{lorkram05}
D.~Lorimer and M.~Kramer, {\em Handbook of Pulsar Astronomy} (Cambridge
  University Press, Cambridge, 2005).

\bibitem{bh86}
D.~C. Backer and R.~W. Hellings, {\em Ann. Rev. Astron. Astrophys.} {\bf 24}
  (1986) 537.

\bibitem{hem06}
G.~B. Hobbs, R.~T. Edwards and R.~N. Manchester, {\em Mon. Not. R. Astron.
  Soc.} {\bf 369}  (2006) 655.

\bibitem{ehm06}
R.~T. Edwards, G.~B. Hobbs and R.~N. Manchester, {\em Mon. Not. R. Astron.
  Soc.} {\bf 372}  (2006) 1549.

\bibitem{ew75}
F.~B. Estabrook and H.~D. Wahlquist, {\em Gen. Relativ. Gravit.} {\bf 6}
  (1975) 439.

\bibitem{ar99}
B.~Allen and J.~D. Romano, {\em Phys. Rev. D} {\bf 59}  (1999) 2001.

\bibitem{m00}
M.~Maggiore, {\em Phys. Rep.} {\bf 331}  (2000) 283.

\bibitem{jb03}
A.~H. Jaffe and D.~C. Backer, {\em Astrophys. J.} {\bf 583}  (2003) 616.

\bibitem{jhs+06}
F.~A. Jenet, G.~B. Hobbs, W.~van Straten, R.~N. Manchester, M.~Bailes, J.~P.~W.
  Verbiest, R.~T. Edwards, A.~W. Hotan, J.~M. Sarkissian and S.~M. Ord, {\em
  Astrophys. J.} {\bf 653}  (2006) 1571.

\bibitem{jhl+05}
F.~A. Jenet, G.~B. Hobbs, K.~J. Lee and R.~N. Manchester, {\em Astrophys. J.
  Lett.} {\bf 625}  (2006) L123.

\bibitem{bd08}
D.~C. Backer and P.~B. Demorest, { Gravitational-wave astronomy with a pulsar
  timing array}, in {\em Frontiers of Astrophysics: A Celebration of NRAO's
  50th Anniversary\/},  eds. A.~H. Bridle, J.~J. Condon and G.~C. Hunt, ASP
  Conference Series, Vol.~395 (2008), pp. 261--270.

\bibitem{dfg+12}
P.~B. Demorest~et al., Limits on the stochastic gravitational wave background
  from the north american nanohertz observatory for gravitational waves
  (2012), arXiv:astro-ph/1201.6641.

\bibitem{rr95}
M.~Rajagopal and R.~W. Romani, {\em Astrophys. J.} {\bf 446}  (1995) 543.

\bibitem{g05}
L.~P. Grishchuk, {\em Phys. U.} {\bf 48}  (2005) 1235.

\bibitem{dv05}
T.~Damour and A.~Vilenkin, {\em Phys. Rev. D} {\bf 71}  (2005) 063510.

\bibitem{sl96}
S.~L. Shemar and A.~G. Lyne, {\em Mon. Not. R. Astron. Soc.} {\bf 282}  (1996)
  677.

\bibitem{klgj03}
A.~Krawczyk, A.~G. Lyne, J.~A. Gil and B.~C. Joshi, {\em Mon. Not. R. Astron.
  Soc.} {\bf 340}  (2003) 1087.

\bibitem{el+11}
C.~M. Espinoza, A.~G. Lyne, B.~W. Stappers and M.~Kramer, {\em Mon. Not. R.
  Astron. Soc.} {\bf 414}  (2011) 1679.

\bibitem{hlk+10}
G.~Hobbs, A.~G. Lyne and M.~Kramer, {\em Mon. Not. R. Astron. Soc.} {\bf 402}
  (2010) 1027.

\bibitem{cs10}
R.~M. Shannon and J.~M. Cordes, {\em Astrophys. J.} {\bf 725}  (2010) 1607.

\bibitem{lhk+10}
A.~Lyne, G.~Hobbs, M.~Kramer, I.~Stairs and B.~Stappers, {\em Science} {\bf
  329}  (2010) 408.

\bibitem{t92}
J.~H. Taylor, {\em Phil. Trans. R. Soc.} {\bf 341}  (1992) 117.

\bibitem{jr06}
B.~C. Joshi and S.~Ramakrishna, {\em Bull. Astron. Soc. India} {\bf 34}  (2006)
  401.

\bibitem{ll94}
A.~G. Lyne and D.~R. Lorimer, {\em Nature} {\bf 369}  (1994) 127.

\bibitem{agm+05}
A.~L. Ahuja, Y.~Gupta, D.~Mitra and A.~K. Kembhavi, {\em Mon. Not. R. Astron.
  Soc.} {\bf 357}  (2005) 1013.

\bibitem{yhc+07a}
X.~P. You~et al., {\em Mon. Not. R. Astron. Soc.} {\bf 378}  (2007) 493.

\bibitem{yhc+07b}
X.~P. You, G.~B. Hobbs, W.~A. Coles, R.~N. Manchester and J.~L. Han, {\em
  Astrophys. J.} {\bf 671}  (2007) 907.

\bibitem{kcs+12}
M.~J. Keith~et al., Measurement and correction of variations in interstellar
  dispersion in high-precision pulsar timing  (2012), arXiv:astro-ph/1211.5887.

\bibitem{mhb+12}
R.~N. Manchester~et al., The parkes pulsar timing array project  (2012),
  arXiv:astro-ph/1210.6130.

\bibitem{kg+12}
U.~Kumar~et al., private comm.

\bibitem{l71}
A.~G. Lyne, {\em Mon. Not. R. Astron. Soc.} {\bf 153}  (1971) 27.

\bibitem{klo+06}
M.~Kramer, A.~G. Lyne, J.~T. O'Brien, C.~A. Jordan and D.~R. Lorimer, {\em
  Science} {\bf 312}  (2006) 549.

\bibitem{crs+12}
F.~Camilo, S.~M. Ransom, S.~Chatterjee, S.~Johnston and P.~Demorest, {\em
  Astrophys. J.} {\bf 746}  (2012) 63.

\bibitem{llm+12}
D.~R. Lorimer, A.~G. Lyne, M.~A. McLaughlin, M.~Kramer, G.~G. Pavlov and
  C.~Chang, {\em Astrophys. J.} {\bf 758}  (2012) 141L.

\bibitem{sjm+12}
M.~P. Surnis, B.~C. Joshi, M.~A. McLaughlin and V.~Gajjar, Discovery of an
  intermittent pulsar: Psr j1839+15  (2012), arXiv:astro-ph/1210.3784.

\bibitem{hlj+11}
R.~van Haasteren~et al., {\em Mon. Not. R. Astron. Soc.} {\bf 414}  (2011)
  3117.

\bibitem{fhb+10}
R.~D. Ferdman, {\em Class. Quantum Grav.} {\bf 27}  (2010) 084014.

\bibitem{ych+11}
D.~R.~B. Yardley~et al., {\em Mon. Not. R. Astron. Soc.} {\bf 414}  (2011)
  1777.

\end{thebibliography}









\end{document}